\documentclass[twocolumn,noshowpacs,preprintnumbers,amsmath,amssymb]{revtex4}
\usepackage{graphicx}
\usepackage{dcolumn}
\usepackage{bm}
\usepackage{hyperref}

\newcommand{\be}{\begin{equation}}
\newcommand{\ee}{\end{equation}}
\newcommand{\bea}{\begin{eqnarray}}
\newcommand{\eea}{\end{eqnarray}}

\newcommand{\nn}{\nonumber}

\newcommand{\beq}{\begin{equation}}
\newcommand{\eeq}{\end{equation}}

\newcommand{\cO}{\mathcal{O}}

\newcommand{\cK}{\mathcal{K}}
\newcommand{\cN}{\mathcal{N}}

\newcommand{\cJ}{\mathcal{J}}
\newcommand{\cR}{\mathcal{R}}

\newcommand{\cV}{\mathcal{V}}

\newcommand{\R}{\text{Re}}

\newcommand{\cref}{{\bf [check ref]}}

\newcommand{\al}{$\alpha^\prime$~}

\newcommand{\tr}{\mathrm{Tr}\:}

\begin{document}
\preprint{MPP-2013-69}
\title{On $\alpha'$ corrections in $\cN=1$ F-theory compactifications
}
\author{Thomas W.~Grimm, Raffaele Savelli and Matthias Wei\ss enbacher}
\affiliation{%
Max-Planck-Institut f\"ur Physik, 
               Munich, Germany 
}
\begin{abstract}
We consider $\cN=1$ F-theory and Type IIB orientifold compactifications 
and derive new $\alpha'$ corrections to the four-dimensional effective action. 
They originate from higher derivative corrections to eleven-dimensional supergravity
and survive the M-theory to F-theory limit. We find a correction
to the K\"ahler moduli depending on a non-trivial intersection 
curve of seven-branes.  
We also analyze a  
four-dimensional higher curvature correction.
\end{abstract}
\maketitle

\section{Introduction}
F-theory is a formulation of Type IIB string theory with seven-branes at varying string coupling \cite{Vafa:1996xn}. It 
captures string coupling dependent corrections in the geometry of an
elliptically fibered higher-dimensional manifold. The general effective actions of 
F-theory compactifications have been studied using the duality with M-theory \cite{Denef:2008wq,Grimm:2010ks}. 
M-theory is accessed through its long wave-length limit provided by eleven-dimensional 
supergravity. This implies that the F-theory effective actions are perturbative in 
the string tension $\alpha'$. Starting with the two-derivative supergravity action 
one derives the classical F-theory effective action. Studying $\alpha'$ corrections 
to this action is of crucial importance for many questions both at the conceptual 
and phenomenological level. In particular, a central task is the analysis of 
moduli stabilization in four-dimensional (4d), $\cN=1$ F-theory compactifications \cite{Denef:2008wq}.

In this work we study a set of $\alpha'$ corrections to 4d, $\cN=1$ F-theory 
effective actions arising from known higher-derivative terms in the 11d supergravity
action. More precisely, we find all corrections induced by a classical Kaluza-Klein reduction 
of the purely gravitational M-theory $R^4$-terms investigated 
in \cite{Green:1997di,Green:1997as,Kiritsis:1997em,Russo:1997mk,
Antoniadis:1997eg,Tseytlin:2000sf} on 
elliptically fibered Calabi-Yau fourfolds. We implement the F-theory limit 
decompactifying the 3d M-theory reduction to four space-time dimensions 
and interpret the resulting corrections in F-theory. Two $\alpha^{\prime 2}$
corrections are shown to survive the limit. We find a correction to the volume of the 
Calabi-Yau fourfold base 
and 
an $R^2$-term in the 4d effective action. Both only depend
on the K\"ahler moduli of the $\cN=1$ reduction. The presence of 
a volume correction in the M-theory reduction on Calabi-Yau fourfolds 
has already been stressed in \cite{Haack:2001jz,Berg:2002es}. 
We find here that, due to this correction, the right 4d, $\cN=1$ coordinates 
are shifted from their classical value. Moreover, the F-theory limit itself, which connects the 3d effective theory to the 4d one, appears to receive corrections as well.
However, when written in terms of the corrected K\"ahler moduli, the 
functional dependence of the K\"ahler metric remains the classical one \footnote{The conclusion that only the K\"ahler coordinates are corrected differs from the statements made in the previous version of this work. It resulted from an incomplete Ansatz for the K\"ahler coordinates. This incompleteness was discovered when performing the reduction of the terms recently obtained in \cite{Liu:2013dna}. The required complete reduction will be presented in detail in \cite{toappear}.}.

It was found 
in \cite{Becker:1996gj} that a general M-theory reduction on a Calabi-Yau fourfold also 
includes a warp factor. In this work we will neglect warping effects. 
There is no warp factor in six dimensions and we comment on the 
$\alpha^{\prime 2}$ corrections in Calabi-Yau threefold reductions of F-theory. 

To give an independent interpretation of these two $\alpha^{\prime 2}$ corrections 
we take the Type IIB weak string coupling limit \cite{Sen:1996vd}. The F-theory 
volume correction is proportional to the volume of the intersection curve of the D7-branes with the 
O7-plane. A simple counting of powers of the string coupling suggests 
that this correction arises from tree-level string amplitudes involving oriented open strings with the topology of a disk, 
and non-orientable closed strings with the topology of a projective plane. This interpretation 
is at odds with the expectation that such a correction arises at open-string one-loop level \cite{Epple:2004ra}. 
This problem appears to be independent of the fact that the correction 
can be absorbed by redefining the K\"ahler coordinates on the moduli space. 
While we will clarify this point further in \cite{toappear}, it would also be crucial, on the one hand, to obtain an independent string derivation of this correction.
Within our approach, on the other hand, one needs to check if there are further corrections in a
fully backreacted M-theory reduction lifted to F-theory that have the same structure as 
the volume correction found here. The 4d higher curvature
correction is matched with a higher curvature modification of the Dirac-Born-Infeld actions 
of D7-branes and O7-planes derived in \cite{Bachas:1999um}. Different $\alpha'$ corrections to F-theory effective actions and their 
weak coupling interpretations have been found in \cite{Grimm:2012rg,GarciaEtxebarria:2012zm}.
A class of $\alpha^{\prime 2}$ corrections in the heterotic string has been discussed recently in \cite{Anguelova:2010ed}.

\section{M-theory reduction}\label{MOrigin}

Our starting point is the long wave-length limit
of M-theory given by 11d supergravity. In particular, we
focus on a well-known higher derivative correction to 
the Einstein-Hilbert term of the form  
\cite{Green:1997di,Green:1997as,Kiritsis:1997em,Russo:1997mk,Antoniadis:1997eg,Tseytlin:2000sf} 
\beq\label{11dAction}
S_{(11)} \supset \frac{1}{(2\pi)^8 l_M^9}\int \ast_{11}{\bf 1}\,\{R^{(11)}_{\rm sc}+\frac{\pi^2 l_M^6}{3^2 2^{11}}\cJ_0\}\,,
\eeq
where $\ast_{D}{\bf 1}={\rm d}^{D}X\sqrt{-G^{(D)}}$ 
is the $D$-dimensional volume element, $R^{(D)}_{\rm sc}$ is 
the $D$-dimensional Ricci scalar, and $l_M$ is
the 11d Planck length.
The correction is given by a Lorentz invariant 
combination of four powers of the Riemann tensor $R^{(11)}$ of the 
schematic form 
\beq  \label{def-cJ0}
  \cJ_0 = t_8t_8 (R^{(11)})^4- \frac{1}{4!} \epsilon_{11} \epsilon_{11} (R^{(11)})^4\ ,  
\eeq
where the precise form of the individual terms is given 
in \eqref{def-t8t8} and \eqref{def-eps11eps11}.
In this work we follow the conventions of \cite{Tseytlin:2000sf}. 
In these 
conventions the metric is dimensionless and only the 
space-time coordinates have dimensions.

If we now compactify this theory on a Calabi-Yau fourfold $Y_4$, the resulting 
3d effective action will include the curvature terms 
(before Weyl rescaling) of the form 
\beq \label{3dEH}
S_{(3)} \supset  \frac{1}{(2\pi)^8 l_M} \int \ast_{3}{\bf 1}\,\big\{ \tilde \cV_4 R_{\rm sc}^{(3)} + l_M^2 \tilde \cV_2 |\cR^{(3)}|^2 \big\}\ ,
\eeq
where $\cR^{(D)}= \frac12 R^{(D)}_{\mu \nu} dx^\mu \wedge dx^\nu$ is the 
curvature two-form in $D$ dimensions, and one has 
\bea \label{abs_def}
   |\cR^{(D)}|^2 \ast_{D}{\bf 1} &=& \tr(\cR^{(D)} \wedge \ast_D \cR^{(D)})  \\
                                            & =& - \frac18 R^{(D)}_{\mu \nu \lambda \rho} R^{(D)\, \mu \nu \lambda \rho}\ast_{D}{\bf 1}\ . \nn
\eea
The volumes appearing in \eqref{3dEH} take the form  
\bea
\tilde \cV_4 &=& \frac{1}{4!}\int_{Y_4} J^4 +\frac{\pi^2}{24} \int_{Y_4} c_3(Y_4) \wedge J \ , \label{def-quantumvolume4} \\
\tilde \cV_2 &=& \frac{\pi^2}{24}  \int_{Y_4} c_2(Y_4) \wedge J \wedge J \ . \label{def-quantumvolume2}
\eea
as shown in appendix \ref{Computation}. All M-theory volumes 
are expressed in units of $l_M$. The two-form $J$ is the K\"ahler form
of $Y_4$, and $c_2(Y_4),\ c_3(Y_4)$ denote the second and third 
Chern class of the tangent bundle of $Y_4$, respectively. 
The quantum volume $\tilde \cV_4$ contains an M-theoretic correction
of order $l_M^6$ to the classical volume $\cV_4$ of the 
internal fourfold $Y_4$ given by the first term in \eqref{def-quantumvolume4}. Both
corrections only depend on the K\"ahler structure of $Y_4$ and do not introduce mixing with 
its complex structure. 
Note that the correction to the 3d Einstein-Hilbert term was already anticipated in \cite{Haack:2001jz,Berg:2002es}.

We close this section by noting that the corrections in 
\eqref{def-quantumvolume4} and \eqref{def-quantumvolume2} are already present in 5d compactifications
of M-theory on a Calabi-Yau threefold $Y_3$. The classical 
volume of the threefold receives a correction proportional 
to the Euler number of the threefold. The four-derivative 
term in \eqref{3dEH} is induced due to a non-trivial integral 
$\int c_2(Y_3) \wedge J$. This term is the supersymmetric 
completion of the mixed gauge-gravitational Chern-Simons term 
$A^\Lambda \wedge \tr(\cR^{(5)} \wedge \cR^{(5)})$, where the vectors $A^\Lambda$ are the 
supersymmetric partners of the K\"ahler structure deformations.

\section{Lift to F-theory} \label{Flift}

We next use the duality between M-theory and F-theory to lift the 
$l_M$-corrections in \eqref{3dEH} to $\alpha'$-corrections of the 4d effective theory
arising from F-theory compactified on $Y_4$. In order to do that, 
we first require that $Y_4$ admits an elliptic fibration over a three-dimensional 
K\"ahler base $B_3$. In this case we can use adjunction formul\ae {} to express Chern classes 
of $Y_4$ in terms of Chern classes of $B_3$. For simplicity, let us restrict to a smooth 
Weierstrass model, i.e.~a geometry without non-Abelian singularities, that 
can be embedded in an ambient fibration with typical fibers being the weighted projective 
space W$\mathbb{P}^2_{231}$. 
This implies having just two types of divisors $D_\Lambda$, $\Lambda=1,\ldots,h^{1,1}(Y_4)$. There 
is the horizontal divisor corresponding to the 0-section $D_0$, and the
vertical divisors $D_\alpha$, $\alpha=1,\ldots,h^{1,1}(B_3)$, corresponding to 
elliptic fibrations over base divisors. Denoting the Poincar\'e-dual two-forms 
to the divisors by $\omega_\Lambda = (\omega_0,\omega_\alpha )$, 
we expand 
\beq
  J=v^0 \omega_0+v^\alpha \omega_\alpha\ ,
\eeq
where $v^0$ is the volume of the elliptic fiber. 
Using adjuction formul\ae {} one derives
\bea
c_3(Y_4)&=&c_3-c_1 c_2-60c_1^3-60\, \omega_0c_1^2\,,\\
c_2(Y_4)&=&c_2+11c_1^2 +12\, \omega_0c_1\,,
\eea
where the $c_i$ on the r.h.s.~of these expressions denote the Chern classes of $B_3$ pulled-back to 
$Y_4$. 

In order to take the F-theory limit of the expression \eqref{3dEH}, 
we need the relation between the 11d Planck length $l_M$ and the string length $l_s$. 
Using M/F-theory duality one obtains 
\beq \label{lM=ls}
2\pi l_s = \tilde{\cV}_4^{1/2} l_M \,.
\eeq
In the F-theory limit 
one sends $v^0 \rightarrow 0$. Such operation decompactifies the fourth dimension 
by sending to infinity the radius 
of the 4d/3d circle in string units: $r \propto \tilde{\cV}_4^{3/2} \rightarrow \infty$.
Henceforth, all 
volumes of the base $B_3$ will be expressed in units 
of $l_s$.

We now have to retain the leading order terms in \eqref{3dEH} in the limit of vanishing fiber 
volume $v^0 \to 0$. We introduce a small parameter $\epsilon$ and express the scaling 
of the dimensionless fields by writing $v^0 \sim \epsilon$. 
As explained in \cite{Grimm:2010ks,Grimm:2011sk} one finds 
$v^\alpha \sim\epsilon^{-1/2} $ and infers the scaling behavior of the classical and quantum volume of $Y_4$ to be $\cV_4 \sim \tilde{\cV}_4 \sim \epsilon^{-1/2}$. 
In the following we use the subscript $_b$ to denote 
quantities of the base that are finite in the limit $\epsilon \to 0$. In particular,
one has 
\beq \label{v=vb}
  2\pi v^\alpha_{b} = \sqrt{v^0} v^\alpha\ ,
\eeq
which holds in the strict $\epsilon \to 0$ limit. Note that $v^\alpha_b$ are 
the volumes of two-cycles of the base in the Einstein frame. 
Inserting \eqref{lM=ls} and \eqref{v=vb}
into \eqref{3dEH}, and neglecting all terms that vanish for $\epsilon$ going to 
zero, we obtain
\beq\label{4dEFT}
S_{(4)}  \supset \frac{1}{(2\pi)^7 l_s^2} \int \ast_{4}{\bf 1}\,\big\{ \tilde \cV^b_3 R_{\rm sc}^{(4)} + l_s^2 \tilde \cV_2^b |\cR^{(4)}|^2 \big\},
\eeq
where 
\bea\label{QuantumBase}
\tilde\cV^b_3 &=&\frac{1}{3!}\int_{B_3} J^3_b-\frac{5 }{8}\int_{B_3} c_1^2(B_3)\wedge J_b\,,\\
\label{V2F}
\tilde \cV_2^b&=&\frac{1}{8}\int_{B_3} c_1(B_3)\wedge J_b^2\,.
\eea
Here $\tilde\cV^b_3$ is now the quantum volume of the base $B_3$.

While \eqref{V2F} enters the 4d effective action \eqref{4dEFT} as a higher derivative correction, \eqref{QuantumBase} contains 
a correction to the classical volume of the base. 
Therefore, the \al correction in \eqref{QuantumBase} would in principle induce a modified F-theory K\"ahler potential. Indeed, starting from the 
M-theory K\"ahler potential, the F-theory limit gives 
\beq\label{MtoF}
K^M=-3\log\tilde\cV_4 \quad \longrightarrow \quad \log R+ K^F\, .
\eeq
Equation \eqref{MtoF} contains the divergent term $\log R$ that is needed 
in the decompactification of the fourth dimension and one identifies $R = 1/r^2$.  
However, to determine $K^F$ explicitly one needs to identify the 
appropriate coordinates to perform the limit \cite{toappear}. 

A first naive guess of coordinates is motivated by 
the reduction of the two-derivative action as in \cite{Grimm:2010ks}
and defines 
$L^\Lambda$ variables as two-cycle volumes normalized by the total quantum volume of the internal 
space, i.e.~one sets
\beq  \label{def-L_o}
     L^\Lambda=\frac{v^\Lambda}{\tilde\cV_4}\ ,\quad   L^0\equiv R\ , \quad L_{b}^\alpha = \frac{v^\alpha_b}{\tilde \cV^b_3}\ .
\eeq
Splitting a factor of $R$ as demanded in \eqref{MtoF} and dropping all terms in  $\tilde\cV_4$
that vanish for $\epsilon \to 0$ one finds
\beq\label{finiteK}
K^F=\log\left[ \left(\frac{1}{3!} L_b^\alpha L_b^\beta L_b^\gamma - \frac{5 }{8 \tilde\cV_b^2}L_b^\alpha k^\beta k^\gamma\right) \cK_{\alpha \beta \gamma}\right]\,.
\eeq
$\cK_{\alpha \beta \gamma}$ is the triple intersection matrix of the base 
and $k^\alpha$ are the expansion coefficients in $c_1(B_3) = k^\alpha \omega_\alpha$. 
One thus would find that a correction quadratic in $k^\alpha$ remains in $K^{\rm F}$.

However, one realizes that one can modify the ansatz  \eqref{def-L_o}
in the presence of the higher curvature correction to 
\beq  \label{def-L_n}
     L^\Lambda=\frac{v^\Lambda}{\tilde\cV_4} \Big(1+ \lambda_1 \frac{\pi^2 \chi(J)}{\tilde\cV_4}\Big) + \lambda_2 \frac{\pi^2 \cK^{\Lambda \Sigma} \chi_\Sigma}{\tilde\cV_4}  \ ,
\eeq
where $\chi_\Lambda = \int_{Y_4} c_3(Y_4) \wedge \omega_\Lambda $, $\chi(J) = \int_{Y_4} c_3(Y_4) \wedge J$, and $\cK^{\Lambda \Sigma}$
is the inverse of $\cK_{\Lambda \Sigma} = \int_{Y_4} \omega_\Lambda \wedge \omega_{\Sigma} \wedge J^2$.  
Remarkably, the two corrections proportional to the constants $\lambda_1,\lambda_2$ 
contain terms that scale with $\epsilon$ precisely as the original $v^\Lambda /\tilde\cV_4$.
Furthermore, one finds that if 
\beq  \label{ab-relation}
    96\, \lambda_1 + 4\, \lambda_2 = 1
\eeq
is satisfied one can write 
\beq \label{KM-simple}
   K^{M} =  \log \left[ \frac{1}{4!}\cK_{\Sigma \Lambda \Gamma \Delta} L^\Sigma L^\Lambda L^\Gamma L^\Delta + \cO(\chi_\Sigma^2) \right]\ , 
\eeq
where we suppressed corrections that are at least quadratic in the $\chi_\Sigma$.
Performing the limit in these $L^\Sigma=(R,L^\alpha_b)$ one finds 
\beq \label{KF-simple}
    K^F=\log\left[ \frac{1}{3!} L_b^\alpha L_b^\beta L_b^\gamma \cK_{\alpha \beta \gamma} + \cO(k^{\alpha\, 4})\right]
\eeq
where the $L^\alpha_b$ are analogously modified by higher curvature
corrections  \cite{toappear}. 

In order to evaluate the K\"ahler metric, the 
precise form of the $\cN=1$ complex 
K\"ahler coordinates is crucial. They can be 
obtained by dimensional reduction of other higher-curvature terms of the 11-dimensional action 
recently obtained in \cite{Liu:2013dna},  as done in \cite{toappear}.
We will comment further on the K\"ahler coordinates
and on their relation to the $L$-variables given in \eqref{def-L_n} in section \ref{Kahlerpot}.

Before giving the Type IIB string interpretation of the \al 
corrections in \eqref{4dEFT}, let us comment on some special 
cases. First of all, when 
the elliptic fibration is trivial, i.e. $Y_4= X_3\times T^2$ with $X_3$ being a 
Calabi-Yau threefold, then 
$c_2(Y_4)=c_2(X_3)$ and $c_3(Y_4)=c_3(X_3)$. Since these have no components along the fiber, all 
corrections in \eqref{3dEH} go to zero and the \al 
corrections in \eqref{4dEFT} are absent in the resulting $\cN=2$ theory. 
Another $\cN=2$ corner of F-theory vacua is reached by taking 
${Y}_4={\rm K3}\times {\rm K3}$, a configuration studied in \cite{GarciaEtxebarria:2012zm} 
with a focus on \al corrections. In this case $c_3({Y}_4)=0$ and the volume 
correction \eqref{QuantumBase} vanishes identically. In contrast, both corrections are 
non-vanishing for 6d, $\cN=1$ vacua arising from F-theory on elliptically fibered 
Calabi-Yau threefolds with classical action derived in \cite{Ferrara:1996wv,Bonetti:2011mw}. The terms are generated by taking the F-theory limit 
of the 5d theory briefly discussed at the end of section \ref{MOrigin}.
Since the threefold volume is 
part of the 6d universal hypermultiplet, the volume correction descends to a modification of the 
hypermultiplet metric of the 4d, $\cN=2$ theory obtained upon further compactification 
on $T^2$. The impact of this correction will, however, crucially depend
on the definition of the $\cN=2$ hypermultiplet coordinates. In summary, comparing all these setups one suspects that the volume 
correction \eqref{QuantumBase} relies on the presence of intersecting 
seven-branes but its significance changes on backgrounds with different number of 
supercharges. This will indeed be confirmed by the analysis of 
section \ref{StringInterpret}.

We close this section with two important remarks. 
Firstly, we stress that there are several additional $l_M$-corrections 
to the fourfold volume surviving the F-theory limit. To see this 
consider the case without seven-branes having a product geometry $Y_4=X_3 \times T^2$.
In this situation corrections involving the Type IIB axio-dilaton $\tau$ have 
been computed by integrating out the whole tower of $T^2$ Kaluza-Klein modes 
of the 11d supergravity multiplet \cite{Green:1997as}. This gives the 
following corrections to the fourfold volume
\beq\label{N=2CY4Volume}
\Delta\cV_4^{\cN=2} \sim \frac{\chi({X}_3)}{(v^0)^{1/2}}E_{3/2}(\tau,\bar\tau)\,,
\eeq
that depends on $\tau$ through the non-holomorphic Eisenstein series $E_{3/2}$ (see also \cite{Collinucci:2009nv})
and has the correct scaling behavior to survive the F-theory limit. 

Our second remark concerns the compactification of 
the 11d action \eqref{11dAction} on ${Y}_4\times S^1$, giving rise to Type IIA 
string theory in two dimensions. The resulting 2d string frame Einstein-Hilbert term takes the form 
\beq
S_{(2)} \supset \frac{1}{(2\pi)^7}\int\ast_{2}{\bf 1}\,g_{\rm IIA}^{-2}\,\tilde\cV^s_4 R\,,
\eeq
where we have used that the length of $S^1$ is $2\pi g_{\rm IIA}^{2/3}l_M=2\pi g_{\rm IIA}l_s$ 
in terms of the Type IIA string coupling. $\tilde\cV^s_4$ denotes the quantum volume of the 
fourfold in units of $l_s$ and it takes the form
\beq\label{IIACY4Volume}
\tilde\cV^s_4 = \cV^s_4-\frac{1}{8}\left(\zeta(3)-g_{\rm IIA}^{2}\frac{\pi^2}{3}\right) \int_{Y_4} c_3(Y_4) \wedge J\,.
\eeq
The above correction contains two pieces already present in the 10d Type IIA 
action as $R^4$ couplings \cite{Green:1997as}. 
The first term in the brackets in \eqref{IIACY4Volume} is tree-level in string perturbation theory and arises from  
integrating out $S^1$ Kaluza-Klein modes analogous to the derivation mentioned for \eqref{N=2CY4Volume}.
The second arises at one-loop of closed strings and is identified with the circle reduction of the volume correction 
in \eqref{3dEH}. We stress that the sign difference in the two contributions arises due to 
their origin in distinct $R^4$ couplings in 10d \cite{Antoniadis:1997eg,Tseytlin:2000sf}. 
The $\zeta(3)$-part can also be derived using mirror symmetry or 
localization techniques as done in \cite{Grimm:2009ef,Honma:2013hma}. 
However, it vanishes in the M-theory limit $g_{\rm IIA}\to\infty$, and hence is of no 
relevance for the present purposes.

\section{String theory interpretation}\label{StringInterpret}

In this section we interpret the corrections in \eqref{4dEFT} in the weak string-coupling limit considered by 
Sen \cite{Sen:1996vd}. This limit is performed in the complex structure moduli space of $Y_4$ and gives a 
weakly coupled description of F-theory in terms of Type IIB string theory on a Calabi-Yau threefold $X_3$ 
with an O7-plane and D7-branes. The Calabi-Yau threefold is a double cover of the 
base $B_3$ branched along the O7-plane. The class of this branching locus is the pull-back of $c_1(B_3)$ to $X_3$. 
When non-Abelian singularities are absent in F-theory, as in the case we consider, the corresponding Sen 
limit contains a single recombined D7-brane wrapping a divisor of class $8c_1(B_3)$, as required by seven-brane 
tadpole cancellation. This D7-brane has the characteristic Whitney-umbrella shape \cite{Collinucci:2008pf,Braun:2008ua}. 

We first discuss the volume correction in \eqref{QuantumBase}. For this correction 
the intersection curve of the D7-brane with the O7-plane plays a crucial role. It is a 
double curve with additional pinch point singularities. However, all we need in the following is
its volume in $X_3$ given by 
\beq  
   \cV_{{\rm D7} \cap {\rm O7}} = 8 \int_{X_3} c^2_1(B_3) \wedge J_b\ ,
\eeq
where we omitted the pullback map from $B_3$ to its double cover $X_3$
in the integrand. Since the intersection numbers of $X_3$ are twice the ones of $B_3$, 
we can immediately read off from \eqref{QuantumBase} the induced correction to the classical volume 
of the Calabi-Yau threefold in units of $l_s$. 
Hence we find in the ten-dimensional Einstein frame the corrected threefold volume 
\beq\label{IIBVolume}
\tilde\cV_3=\cV_3-\frac{5}{64}\, \cV_{{\rm D7}\cap {\rm O7}}\,,
\eeq
where $\cV_3$ is the classical volume of $X_3$ that is twice the classical volume of $B_3$.
Note that the quantum correction in \eqref{IIBVolume}
can alternatively be expressed in terms of the volume of the self-intersection curve 
of the O7-plane by using tadpole cancellation. We stress that the correction is 
of order $\alpha^{\prime2}$ since two of the original six derivatives 
in M-theory have been absorbed by the integration on the elliptic fiber.

It is worth noting that the non-triviality 
of the elliptic fibration causes the appearance of an \al correction already at order two, a 
phenomenon also observed in \cite{GarciaEtxebarria:2012zm} for a different correction in F-theory 
compactifications on K3$\times$K3. 

In order to give the string theory interpretation of the 
correction in \eqref{IIBVolume} we have to identify the string amplitude 
capturing it. We first look at the 4d effective action in the string frame
with Einstein-Hilbert term 
\beq\label{StringFrameIIB}
S_{(4)}\supset\frac{1}{(2\pi)^7l_s^2}\int\ast^s_{4}{\bf 1}\left\{\frac{\cV^s_3}{g_{\rm IIB}^{2}}-\frac{5   \cV_{{\rm D7} \cap {\rm O7}}^s }{64\, g_{\rm IIB}}\right\}R^s_{\rm sc}\,, 
\eeq
where the superscript $^s$ denotes quantities computed using the string frame metric. 
Let us indicate which string amplitude might generate the correction in \eqref{StringFrameIIB}.
Recall that the power of the string coupling constant in a given amplitude coincides with $-\chi(\Sigma)$
modified by the number of insertions of vertex operators on the string world-sheet $\Sigma$. Both contributions in \eqref{StringFrameIIB} are expected to 
arise from amplitudes with two graviton insertions and we study the relative $g_s$-power of the 
two terms. 
The general formula for the Euler number of Riemann surfaces, possibly non-orientable and with boundaries, is
\beq
\chi(\Sigma)=2-2g-b-c\,,
\eeq
where $g,b,c$ denote the genus, the number of boundaries, and the number of cross caps, respectively. 
Therefore, we immediately see that the volume correction in \eqref{IIBVolume} should arise from a string 
amplitude that involves the sum over two topologies: The disk ($g=c=0, b=1$) and the projective 
plane ($g=b=0, c=1$). They correspond to the tree-level of orientable open strings and 
non-orientable closed strings, respectively.  This interpretation seemingly contradicts the 
expectation that such a correction arises at open-string one-loop level \cite{Epple:2004ra}. 
We hope to clarify 
this point further in future work. 
Let us stress that 
we also cannot exclude the possibility that there are further corrections in a
fully backreacted M-theory reduction lifted to F-theory that have the same $\alpha'$-order and field dependence 
as the volume correction found here. 

Let us next give a string theory interpretation of the 4d higher derivative correction in \eqref{4dEFT}. 
In fact, at weak string coupling, the coefficient \eqref{V2F} can be written as
\beq \label{O7D7-split}
\tilde\cV^{\rm IIB}_2=\frac{1}{96} \big(\cV_{\rm D7} + 4 \cV_{\rm O7} \big)\,,
\eeq
where $\cV_{\rm D7}$ and $\cV_{\rm O7}$ are the volumes of the D7-brane and the O7-plane in $X_3$, respectively.
Both volumes are in the Einstein frame and in units of $l_s$. By tadpole cancellation one has $\cV_{\rm D7}=8 \cV_{\rm O7}$.
However,  in \eqref{O7D7-split} we have split the volumes according to the appearance of the corresponding divisors in 
the F-theory discriminant. The relative factor in the volume split is in agreement 
with the relative factor in the higher curvature terms of the Chern-Simons actions of D7-branes and O7-planes. 
These have been studied to derive the 4d higher curvature term proportional to $\tr (\cR^{(4)} \wedge \cR^{(4)})$
in \cite{Grimm:2012yq}, which is the supersymmetric partner of the $\tr (\cR^{(4)} \wedge \ast_4 \cR^{(4)})$ term in \eqref{4dEFT}.
Translated to the string frame the higher derivative correction in \eqref{4dEFT} becomes
\beq
S_{(4)}\supset\frac{1}{96 \, (2\pi)^7  }\int\ast^s_{4}{\bf 1}\,\frac{\big(\cV^s_{\rm D7} + 4 \cV^s_{\rm O7} \big)}{g_{\rm IIB}}|\cR^{(4)}_s|^2\,,
\eeq
where we see that this correction has the same string loop order as the one in \eqref{StringFrameIIB}.
This term is expected to directly arise from a higher curvature correction of the string-tree-level Dirac-Born-Infeld action on 
the D7-brane and O7-plane as discussed in \cite{Bachas:1999um} (see, in particular, equation (3.5)). 

\section{Remarks on the K\"ahler potential and Type IIB vacua} \label{Kahlerpot}

In this final section we comment on the structure of the 
4d, $\cN=1$ K\"ahler potential $K^F$ given in \eqref{KF-simple} and 
analyze its properties. In order to derive the kinetic terms of 
the moduli and the scalar potential, one first needs to express $K^F$
in terms of the $h^{1,1}(B_3)$ correct $\cN=1$ complex coordinates $T_\alpha$.
Starting in M-theory one has $h^{1,1}(Y_4)$ complex coordinates $T_\Lambda$.
Classically the real parts of $T_\Lambda$ are the volumes of the 
divisors of $Y_4$, while the imaginary part is the integral of the 
M-theory six-form over the same divisors. Including higher curvature
corrections also the $T_\Lambda$ might be shifted as \footnote{The crucial second term was not considered in the 
previous version of this work. Together, both shifts allow to revise the interpretation of the 
correction.}
\beq\label{Tmoduli}
{\rm Re}\, T_{\Lambda} = \frac{1}{3!} K_{\Lambda} \Big(1+ \kappa_1 \frac{\pi^2}{\tilde \cV_4} \chi(J) \Big)    + \kappa_2 \pi^2 \, \chi_\Lambda\,,
\eeq
where we abbreviated $K_\Lambda = \int_{Y_4}  J^3 \wedge  \omega_\Lambda$, and used $\chi_\Lambda$, $\chi(J)$
defined after \eqref{def-L_n}. All volumes are expressed in units of $l_M$.
The shifts proportional to the constants $\kappa_1,\kappa_2$ 
could be expected in analogy to the proposal of \cite{Grimm:2010ks} that 
the ${\rm Re} T_{\Lambda} $ are related to the periods of the mirror Calabi-Yau fourfold. 
To derive corrections to the 
$T_\Lambda$ the modification of their kinetic terms has to be computed \cite{toappear}. 
For the correction considered here one would have to dimensionally reduce 
the 11d higher curvature term in \eqref{11dAction} including fluctuations 
of the Calabi-Yau metric. Remarkably, it turns out in \cite{toappear} that the correct 
choice of $T_\Sigma$ is such that the Legendre dual variables to 
${\rm Re}\, T_{\Lambda}$ are  
\beq \label{L-Legendre}
   L^\Sigma = -  \frac{\partial K^M}{ \partial {\rm Re}\, T_\Sigma } = - 2 K^M_{T_\Sigma}\ ,
\eeq
with $K^M$ as in  \eqref{KM-simple} and $L^\Sigma$ of the form \eqref{def-L_n} with 
\beq\label{correctCoeff}
   \lambda_1 = \frac{1}{3 \cdot 24},\ \lambda_2 = -\frac{1}{12},\  \kappa_1 = \frac{1}{24},\ \kappa_2 = -\frac{1}{24} \ ,
\eeq
Hence \eqref{ab-relation} is satisfied and ${\rm Re}\, T_\Lambda L^\Lambda = 4$. This in turn implies that at linear order in $\chi_\Sigma$ the no-scale-like property 
of the corrected M-theory K\"ahler potential is still satisfied
\beq \label{no-scale1}
 K^M_{T_\Lambda} K^{M\, T_\Lambda \bar T_{\Sigma}} K^M_{\bar T_\Sigma} = 4 \,,
\eeq
where $K^{T_\Lambda \bar T_{\Sigma}}$ is the inverse K\"ahler metric.
The result \eqref{no-scale1} can be attributed to the fact that 
${\rm Re}\, T_\Sigma$ and $L^\Sigma$ are Legendre dual 
variables via \eqref{L-Legendre} and $K^M$ takes  the 
simple form \eqref{KM-simple}. 
This is in contrast to the claim made in the 
previous version of this work and can be traced 
back to having not considered a 
sufficiently general ansatz for $L^\Sigma,T_\Sigma$.
Clearly, the correct choice for $L^\Sigma,T_\Sigma$
can only be evaluated by a more complete reduction 
as done in \cite{toappear} in which we discovered the 
incompleteness in the original ansatz for \eqref{Tmoduli}.

In order to express the M-theory K\"ahler potential in terms of the $T_\Lambda$-moduli, 
one would have to solve \eqref{Tmoduli} for $L^\Lambda$, and insert the solution back into \eqref{MtoF}. 
Strictly speaking one would next have to perform a Legendre transformation replacing 
$\R T_0$ with $R$ and work with the modified kinetic potential $\tilde K(R,T^b_\alpha)$, where $T_\alpha^b$ are the 4d $\cN=1$ coordinates. 
However, in the F-theory limit one finds the expression \eqref{MtoF}, with $K^F$ given in \eqref{KF-simple}, which at order $k^{\alpha\,2}$  still satisfies the no-scale property  as 
\beq
 K^F_{T^b_\alpha} K^{F\, T^b_\alpha \bar T^b_{\alpha}} K^F_{\bar T^b_\alpha} = 3 \,.
\eeq
Therefore, as was the case in 3d, also in 4d the quantum correction only affects the K\"ahler coordinates, whereas the K\"ahler metric remains classical. 
It turns out in \cite{toappear} that in order to derive the 4d quantum K\"ahler coordinates from the 3d ones given in \eqref{Tmoduli}
one can also include $\alpha'$ corrections to the F-theory limit itself. This freedom allows to bring ${\rm Re} T_\alpha^b$ into 
the form 
\beq\label{Tbmoduli}
\frac{{\rm Re} T_\alpha^b}{
 (2\pi)^2}  = \frac{1}{2}\cK^b_\alpha  - \frac{5}{16 \tilde \cV_3^b} c\, \cK^b_\alpha  \chi^b(J_b)  + \frac{5}{8} (3c-2) \chi_\alpha^b\ ,
\eeq
where $c$ is a constant that parametrizes the freedom to modify the F-theory limit. 
Here we have defined $\cK^b_\alpha = \cK_{\alpha \beta \gamma} v^\beta_b v^\gamma_b$, 
$\chi_\alpha^b = \int_{B_3} c_1^2(B_3) \wedge \omega_\alpha $, and $\chi^b(J_b) = \chi_\alpha^b v^\alpha_b$. The 
${\rm Re}\, T^b_{\alpha}$ as in \eqref{Tbmoduli} are related to the $L^\alpha_b$ variables via Legendre duality with $K^F$ as in \eqref{KF-simple}. 
Note that when $c=0$ the result \eqref{Tbmoduli} for the corrected ${\rm Re}\, T^b_\alpha$ contains just a constant shift 
from the classical value.

Before concluding, it is worth remarking that the fourfold volume \eqref{def-quantumvolume4} will in principle get further corrections beyond linear order in $\chi_{\Sigma}$. Likewise, the system of Legendre dual coordinates will  be modified and $L^\Sigma, {\rm Re}\,T_\Sigma$ will have more general expressions reducing to \eqref{def-L_n}, \eqref{Tmoduli} at linear order in $\chi_{\Sigma}$. However, since we have no control over them from a direct derivation of the moduli kinetic term, we are not able to determine whether the K\"ahler metric will keep its classical form and the no-scale property will still be satisfied beyond linear order in $\chi_{\Sigma}$.

\acknowledgements
		
We like to thank Ralph Blumenhagen, Federico Bonetti, Andr\'es Collinucci, Ioannis Florakis, 
Michael Haack, Dieter L\"ust, Francisco Pedro, Markus  Rummel, Stephan Stieberger, and Alexander Westphal 
for useful discussions and comments.

\appendix

\section{Computation}\label{Computation}

We use the conventions of  \cite{Nakahara} for the definition 
of the Riemann tensor and related quantities. The background geometry is the product manifold 
$M_{11}=\mathbb{R}^{1,2} \times Y_4$, where the flat space has signature $\{-,+,+ \}$ 
and $Y_4$ is the Calabi-Yau fourfold.  
External indices are denoted by $\mu, \mu' $. For the coordinates on $Y_4$ we use real and complex
indices denoted by $a, a'$
and $\alpha, \beta, \gamma, \delta$, respectively. Indices of the coordinates of the total space $M_{11}$
will be written in capital Latin letters $N, N'$. Furthermore, the convention for the totally 
anti-symmetric tensor in Lorentzian space in an orthonormal frame is $\epsilon_{012...10} = \epsilon_{012}=+1$.

 The curvature two-form for Hermitian manifolds is defined as
 \begin{equation}
 {\cR^{\alpha}}_{\beta} =  {{R^{\alpha}}_\beta}_{\gamma \bar\delta} dz^\gamma \wedge d\bar{z}^{\bar\delta}\,, 
 \end{equation}
 and one has
 \bea
  \tr{\cR}  \! &=& \! {{R^\alpha}_\alpha}_{\gamma \bar\delta}dz^\gamma \wedge d\bar{z}^{\bar\delta}\,, \\ 
 \tr{\cR^2} \! &=& \! {{R^{\alpha}}_{\beta}}_{\gamma \bar\delta} {{R^{\beta}}_{\alpha}}_{\gamma_1 \bar\delta_1}
                        dz^{\gamma} \wedge d\bar{z}^{\bar\delta}\wedge dz^{\gamma_1} \wedge d\bar{z}^{\bar\delta_1}\,,\nn   \\
 \tr{\cR^3} \! &=& \! {{R^{\alpha}}_{\beta}}_{\gamma \bar\delta} {{R^{\beta}}_{\beta_1}}_{\gamma_1 \bar\delta_1} {{R^{\beta_1}}_{\alpha}}_{\gamma_2 \bar\delta_2}
                       dz^{\gamma} \wedge d\bar{z}^{\bar\delta} \dots   d\bar{z}^{\bar\delta_2}\,. \nn
 \eea

The correction to the 11d Einstein-Hilbert term in \eqref{11dAction} is given by $\cJ_0$
schematically defined in \eqref{def-cJ0} to be
\beq \label{def-cJ0app}
\mathcal{J}_0 = t_8t_8 R^4 - \frac{1}{4!} \epsilon_{11} \epsilon_{11} R^4\, . 
\eeq
Following \cite{Freeman:1986zh} we define in real coordinates
 \beq
 \begin{array}{ll}
  t^{a_1 a_2 \dots a_8}_8  =
  & - 2 \left(  \delta^{\lceil a_1[a_3| }\delta^{ |a_2 \rceil a_4] }\delta^{\lfloor a_5 \langle a_7| }\delta^{|a_6\rfloor a_8 \rangle } \right. \\
 &  + \delta^{\lceil a_1[a_5| }\delta^{|a_2\rceil a_6] }\delta^{\lfloor a_3 \langle a_7| }\delta^{| a_4\rfloor   a_8 \rangle }\\
 & + \left. \delta^{\lceil a_1[a_7| }\delta^{|a_2\rceil a_8] }\delta^{\lfloor a_3 \langle a_5| }\delta^{| a_4\rfloor a_6 \rangle }  \right) \\ [.1cm] 
  &   +  8 \left(  \delta^{\langle a_2|  [ a_3   }\delta^{ a_4]  \lceil a_5  }\delta^{ a_6\rceil   \lfloor a_7  }\delta^{ a_8\rfloor | a_1\rangle  } \right. \\  
  &  +  \delta^{\langle a_2|  [ a_5   }\delta^{ a_6]  \lceil a_3  }\delta^{ a_4\rceil   \lfloor a_7  }\delta^{ a_8\rfloor | a_1\rangle  }   \\ 
  &  \left.+  \delta^{\langle a_2|  [ a_5   }\delta^{ a_6]  \lceil a_7  }\delta^{ a_8\rceil   \lfloor a_3  }\delta^{ a_4\rfloor | a_1\rangle  }  \right)\, .
  \end{array}
   \eeq
 The symbols $[ \; ] ,  \lceil  \; \rceil, \lfloor \;  \rfloor, \langle  \;\rangle$ denote anti-symmetrization
 and indices in between two vertical lines $|$ are omitted in the respective bracket. 
 The expression is anti-symmetrized in the following pairs 
 of indices $(a_1 a_2), (a_3 a_4), (a_5 a_6), (a_7 a_8)$ respectively.
 Thus we have for the term $t_8 t_8 R^4$ in real coordinates
 \beq \label{def-t8t8}
 t_8 t_8 R^4 = t_8{_{a_1 \dots a_8}} t_8^{a'_1 \dots a'_8} {R^{a_1 a_2}}_{a'_1 a'_2} \cdots {R^{a_7 a_8}}_{a'_7 a'_8}\,.
 \eeq
The second term in \eqref{def-cJ0app} can be written as 
\beq \label{def-eps11eps11}
  \frac{1}{4!}  \epsilon_{11}\epsilon_{11}R^4 = \frac14 E_8(M_{11})\, ,
\eeq 
where one uses the general definition in real coordinates 
\begin{align}
 \label{eq:DefEn}
&E_n(M_D) = \frac{1}{(D-n)!}\epsilon_{N_1\cdots N_{D}}\epsilon^{N_1\cdots N_{D-n}N'_{D-n+1}\cdots N'_{D}} \nn \\
& {R^{N_{D-n+1}N_{D-n+2}}}_{N'_{D-n+1}N'_{D-n+2}} \cdots {R^{N_{D-1}N_{D}}}_{N'_{D-1}N'_{D}}\,,  \nn \\
\end{align}
where $n >0$ and $D$ being the real dimension of the manifold $M_D$. 
 
The Chern classes can be expressed in terms of the curvature two-form $\cR$ as
 \bea
 c_1 &=& i \tr{\cR}\, , \quad c_2 =  \frac{1}{2!}\left(  \tr{\cR^2} - (\tr{\cR})^2\right)\, , \\  
  c_3 &=& \frac{1}{3}c_1 c_2 + \frac{1}{3}c_1\wedge  \tr{\cR^2} - \frac{i}{3} \tr{\cR^3} \,  \nn \\
  c_4 &=&  \frac{1}{24}\left( c_1^4   - 6 \, c_1^2\, \tr{\cR^2}  - 8\,i \, c_1 \, \tr{\cR^3} \right) \,  \nn  \\
  &+ & \frac{1}{8}((\tr{\cR^2})^2-2 \, \tr{\cR^4})\, . \nn 
 \eea
The  Chern classes of a Calabi-Yau fourfold reduce 
to $c_3(Y_4) = -\frac{i}{3}\,\tr{\cR^3}$ and $c_4 =  \frac{1}{8}((\tr{\cR^2})^2-2 \tr{\cR^4})$. 

Let us first compute $E_8(M_3 \times M_8)$ 
for a generic product space.
By using the definition \eqref{eq:DefEn}, splitting indices and applying Schouten identities it is straightforward to show that 
\beq \label{E8_splitt_app}
E_8 (M_3 \times M_8) = - E_8(M_8) + 4 \, E_2(M_3) E_6(M_8)\, ,
\eeq
where $ E_2(M_3) = -2 R^{(3)}_{\rm sc}$ and
 \beq
 \label{eq:E6M8}
 E_6(M_8) =   6!  {R^{ [a_1 a_2}}_{ a_1 a_2} \cdots {R^{a_{5}a_{6}]}}_{a_{5} a_{6}}.
 \eeq

Next we evaluate $c_3 \wedge J$ on $Y_4$, where $J= i  g_{\alpha \bar\beta} dz^{\alpha} \wedge d\bar{z}^{\bar\beta}$ is the K\"ahler form.
Expressing $c_3$ in holomorphic coordinates we find
\beq
\label{eq:c3J}
c_3\wedge J =  2  {R^{\alpha_0}_{\phantom{\beta_0} \beta_0 \gamma_0}}^{ [\gamma_0}_{\phantom{\beta_0}} {R^{|\beta_0|}_{\phantom{|\beta_0|} \beta_1 \gamma_1}}^{\gamma_1} 
                  { {R^{|\beta_1|}}_{\alpha_0 \gamma_2}} ^{\gamma_2]} \ast_{8} {\bf 1} \,.
\eeq

In order to compare \eqref{eq:E6M8} and \eqref{eq:c3J} one has to change from real coordinates to complex ones. For $M_8=Y_4$ one finds
\bea \label{E6Y4}
  \begin{array}{l}
 E_6(Y_4)\ast_{8} {\bf 1}   =     128 \left(  {{{R^{\alpha_0}}_{\beta_0}}_{\gamma_0}}^{ \delta_0} {{{R^{\beta_0}}_{\alpha_1}}_{\delta_0}}^{\delta_1}{ {{R^{\alpha_1}}_{\alpha_0}}_{\delta_1}} ^{\gamma_0}    \right.
 \\ \\
 + \left.     {{{R^{\alpha_0}}_{\beta_0}}_{\gamma_0}}^{ \delta_0} {{{R^{\beta_0}}_{\alpha_1}}_{\delta_1}} ^{\gamma_0}  { {{R^{\alpha_1}}_{\alpha_0}}_{\delta_0}}^{\delta_1}  \right)\ast_{8} {\bf 1} \,.
 \end{array}
 \eea
Comparing  \eqref{eq:c3J} and \eqref{E6Y4} we find
\beq
E_6(Y_4) \ast_{8} {\bf 1} =  3 \cdot 2^{7}  \;\; c_3\wedge J\,.
\eeq
Additionally we have
\beq \label{E8_int_eval}
\frac{1}{4}E_8(Y_4) \ast_{8} {\bf 1} = 1536 \; c_4\ .
\eeq

Finally we find for the reduction of $t_8t_8 R^4$ the terms
\bea \label{t8exp}
t_8t_8 R^4 \ast_8 {\bf 1}= - 96 {(R^{\mu} _{\,\,\nu})}_{\mu' \nu'}  {(R^{\nu} _{\,\, \mu})}^{\mu' \nu'} (R^{\alpha}_{\,\, \beta})_{\gamma_0}^{\,\,\,\delta_1} \nn \\  [.1cm] (R^{\beta}_{\,\, \alpha})_{\delta_1}^{\,\, \gamma_0} \ast_8 {\bf 1}  
 + \,1536 \, c_4 \, +  \ldots  \, ,
\eea
where the dots indicate purely external terms.
 To rewrite this result let us consider the following terms
 \bea
 c_2 \wedge J^2  &=&   - 2\,  {\delta^{[\gamma_0}}_{\delta_0}  {\delta^{\gamma_1]}}_{\delta_1} 
 {{({R^{\alpha}}_{\beta})}_{\gamma_0}}^{\delta_0} {{({R^{\beta}}_{\alpha})}_{\gamma_1}}^{\delta_1}  \ast_{8} {\bf 1} \nn \\ 
 &=&  - {{({R^{\alpha}}_{\beta})}_{\gamma_0}}^{\delta_1} {{({R^{\beta}}_{\alpha})}_{\delta_1}}^{\gamma_0} \ast_{8} {\bf 1}\, .
 \eea
Furthermore, we find
\beq
\tr{\left[  \cR^{(3)} \wedge \ast_{3} \cR^{(3)}\right]} = \frac{1}{8} {({R^{\mu}} _\nu)}_{\mu' \nu'}  {({R^{\nu}} _\mu)}^{\mu' \nu'}  \ast_{3} {\bf 1}\, ,
\eeq
with $\cR^{(3)}$ being the real curvature two-form as in \eqref{abs_def}.
Hence we conclude
\bea \label{t8t8eval}
t_8t_8 R^4 \ast_{11} {\bf 1} &=&  3 \cdot 2^8 \, \tr{\left[  \cR^{(3)} \wedge \ast_{3} \cR^{(3)}\right]}  \wedge   \left( c_2 \wedge J^2  \right) \nn \\ 
&& + \,1536 \, c_4 \,  + \ldots\, ,
\eea
where we have omitted the same purely external terms as in \eqref{t8exp}.

To conclude we use \eqref{def-cJ0app}, \eqref{E8_splitt_app}, \eqref{E8_int_eval} and \eqref{t8t8eval} to 
find for the internal terms of $\cJ_0$
\beq \label{cJ_int}
 (\cJ_0)_{\rm int} \ast_8 {\bf 1}= \left[(t_8t_8 R^4)_{\rm int} +\frac{1}{4}E_8(Y_4)\right]\ast_8 {\bf 1} = 3072 \, c_4 \, , 
\eeq
which integrates to $3072 \, \chi$ on $Y_4$. The linear combination 
with a different relative sign in
equation \eqref{cJ_int} obviously vanishes on $Y_4$.  This is of physical importance 
as discussed e.g.~in \cite{Gross:1986iv}.

\end{document}